\documentclass[12pt]{article}
\usepackage{graphicx}
\usepackage{graphicx}
\usepackage{amsfonts}
\usepackage{amssymb}
\usepackage{amsmath}

\newlength{\dinwidth}
\newlength{\dinmargin}
\def\lapproxeq{\lower .7ex\hbox{$\;\stackrel{\textstyle
<}{\sim}\;$}}
\def\gapproxeq{\lower .7ex\hbox{$\;\stackrel{\textstyle
>}{\sim}\;$}}

\def\bea{\begin{eqnarray}}
\def\eea{\end{eqnarray}}
\def\Qtbold{\mbox{\boldmath${Q}$}_t}
\def\qbold{\mbox{\boldmath${q}$}}

\begin{document}
\titlepage
\vspace*{1in}
\begin{center}
{\Large \bf Unintegrated gluon distributions in a photon from the CCFM equation
in the single loop approximation}\\
\vspace*{0.4in}
Agnieszka Gawron and Jan   Kwieci\'nski
\\
\vspace*{0.5cm}

{\it H. Niewodnicza\'nski Institute of Nuclear Physics,
 Krak\'ow, Poland} \\
\end{center}
\vspace*{1cm}
\centerline{(\today)}

\vskip1cm
\begin{abstract}
The system of CCFM equations for unintegrated parton distributions in a photon is considered in the single loop
approximation.  We include quarks and non-singular parts of the splitting functions
in the corresponding evolution equations.  We solve the system of  CCFM equations
utilising the transverse coordinate representation which diagonalises these equations
in the single loop approximation.  The results for the unintegrated gluon
distributions in a photon are presented and  confronted with the approximate
form expressing those distributions in terms of the integrated gluon and quark 
distributions and a suitably defined Sudakov-like form factor.
\end{abstract}

\section{Introduction}

   Inclusive quantities describing the hard processes  are controlled  in the
QCD
improved parton model by the scale
dependent quark and gluon distributions which depend upon the longitudinal momentum fraction
$x$ and upon the hard scale $Q^2$.  In order to describe  less inclusive quantities
which are sensitive to the transverse momentum of the parton it is however
necessary to consider the  distributions unintegrated over the transverse
momentum of the parton \cite{DDT}-\cite{LUNDSMX}.
Those unintegrated distributions are described in perturbative QCD by the Ciafaloni-Catani-Fiorani-
Marchesini (CCFM) equation \cite{CCFM},\cite{GM1} based upon quantum coherence which implies
angular ordering \cite{DKTM}. It embodies in a unified way
the (LO) DGLAP evolution
and BFKL dynamics at low $x$.\\

  Existing  analyses
of the CCFM equation concern predominantly parton distributions in a nucleon \cite{LUNDSMX}, \cite{BRW} -
\cite{JUNGS}.  The purpose of this
paper is to extend this analysis to the case of the unintegrated  parton distributions in
a photon.
We limit ourselves to the so called 'single loop' approximation in which the CCFM
equation is equivalent to the LO DGLAP evolution \cite{BRW,GMBRW}.   We shall
utilise the fact that in
this approximation the CCFM equation is diagonalised by the Fourier-Bessel transform and
so
one can explore the transverse coordinate representation of this equation \cite{JK}.
The transverse coordinate representation conjugate to the transverse momentum of the
parton   has proved to be very useful in studying $p_t$ distributions within
the DGLAP framework and it   has  been widely explored in the analysis of the soft gluon resummation
effects
in $e^+ e^-$ collisions \cite{BASSETTO},\cite{KODAIRA}, in the $p_t$ distribution
of  Drell-Yan pairs \cite{PT} etc.  The formalism of transverse  coordinate
representation adopted in our analysis of the CCFM equation is  similar
to that used in those studies.\\

 The single-loop approximation  of the CCFM equation which we shall
use  neglects important small $x$ effects and so it
may not be reliable at (very) small $x$.  It should however become an
adequate approximation
at moderately
small values of $x$ (i.e. $x >0.01$ or so) which is relevant phenomenologically e.g.
for the description of the heavy quark production in $\gamma \gamma$ collisions at presently
available energies \cite{SZCZUREK}.\\

  The CCFM equation is usually considered only for the gluonic
sector and, in principle, with only the singular parts of the $g\rightarrow gg$
splitting
functions included in the evolution.  In order to have a formalism which is  phenomenologically
relevant
at large and moderately small values of $x$ one has to incorporate also
the quark distributions and  the complete splitting functions.  This is straigtforward in
the 'single loop' approximation which, after integration over  the transverse momentum
of the partons, should reduce  the CCFM equations to the conventional DGLAP
evolution equations.\\

  The content of our paper is as follows:
In the next Section we introduce the system of CCFM equations in the single loop approximation
for the unintegrated parton distributions in a photon.
In Section 3 we discuss the transverse coordinate representation   which partially
diagonalises the system of CCFM equations.  In Section 4 we present results
of the numerical solution of the CCFM equation(s) for the unintegrated gluon distributions in a
photon.  We do also discuss
approximate treatment of these equations which allows to relate the unintegrated
gluon distributions in a photon to the integrated gluon and quark distributions and
the suitably defined Sudakov-like form-factor.
 Finally, in Section 5,  we summarise our main results and give our conclusions.\\

\section{The CCFM equation in the single loop approximation for the parton distributions
in a photon}

 In this Section we introduce the system of CCFM equations for the unintegrated
parton distributions in a photon.  We extend  the CCFM framework by including the
quark distributions and the non-singular parts of the splitting functions.
We  limit ourselves to the single-loop approximation which should be adequate
in the region of moderately small values of $x$.  \\

  The  original Catani, Ciafaloni, Fiorani, Marchesini
(CCFM) equation \cite{CCFM} for the unintegrated, scale dependent gluon distribution
$f_g(x,Q_t,Q)$ which is generated by the sum of ladder diagrams with  angular
ordering along the chain  has the following form:

\vspace*{1cm}

\begin{gather}
f_g(x,Q_t,Q)=\tilde f_g^0(x,Q_t,Q) + \int{d^2\qbold \over \pi q^2}
\int_x^1 {dz\over z} \Theta(Q-qz) \Theta(q-q_0) { \alpha_s\over
2\pi } \Delta_s (Q,q,z)\times
\nonumber\\*[5mm]
\times\left[2N_c\Delta_{NS}(Q_t,q,z) + {2N_cz\over (1-z)}
 f\left({x\over z},|\Qtbold+(1-z)\qbold|,q\right)\right] ,
\label{ccfm1}
\end{gather}
where $\Delta _S(Q,q,z)$ and $\Delta_{NS}(Q_t,q,z)$ are the Sudakov and non-Sudakov
form factors.  They are given by the following expressions:
\vspace*{5mm}
\begin{equation}
\Delta _S(Q,q,z)=exp\left[-\int_{(qz)^2}^{Q^2}{dp^2\over
p^2}{\alpha_s\over 2 \pi} \int _0^{1-q_0/p}dzzP_{gg}(z)\right]
,
\label{ds}
\end{equation}

\begin{equation}
\Delta_{NS}(Q_t,q,z)= exp\left[-\int_z^1{dz'\over z'}
\int_{(qz')^2}^{Q_t^2} {dp^2\over p^2}{2N_c\alpha_s\over 2
\pi}\right] .
\label{dns}
\end{equation}
The variables $x,Q_t,Q$ denote the longitudinal momentum fraction,
transverse momentum of the gluon and the hard scale respectively. The
latter is defined in terms of the maximal emission angle \cite{LUNDSMX,CCFM}.
The constraint $\Theta(Q-qz)$ in equation (\ref{ccfm1}) reflects the angular ordering and the inhomogeneous term
$\tilde f^0(x,Q_t,Q)$ is related to the input non-perturbative gluon distribution.
It also contains effects of both the Sudakov and non-Sudakov form-factors \cite{KMSU}. \\

In order to make the CCFM formalism realistic in the region of large and moderately
small values of $x$ we should  introduce, besides the unintegrated
gluon distribution $f_{g}(x,Q_t,Q)$ also the unintegrated quark  distributions
$f_{q_i}(x,Q_t,Q)$, where $i$ numerates the quark flavour, 
 and  include the $q\rightarrow gq$, ($\bar q \rightarrow g\bar q$) and $g \rightarrow 
\bar q q$ transitions along the chain.  In order to get exact
correspondence with the complete LO DGLAP evolution one should also use  complete
splitting functions and not only their singular components. In the region of large and moderately
small values of $x$
one can   introduce the 'single loop'
approximation which corresponds to the replacement of  the angular ordering
constraint
$\Theta(Q-qz)$  by $\Theta(Q-q)$ and to setting the non-Sudakov form-factor
$\Delta_{NS}$  equal to  unity \cite{BRW,GMBRW}.\\

   It is convenient to consider the unintegrated singlet {$S$}
and non-singlet ($NS$) quark distributions:
\begin{equation}
f_S(x,Q_t,Q)=2\Sigma_{i=1}^f f_{q_i}(x,,Q_t,Q) ,
\label{singlet}
\end{equation}

\begin{equation}
f_{NS}(x,Q_t,Q)=2\Sigma_{i=1}^f e_i^2
f_{q_i}(x,,Q_t,Q)-<e^2>f_S(x,Q_t,Q) ,
\label{nonsinglet}
\end{equation}

where

\begin{equation}
<e^k> = {1\over f} \Sigma_{i=1}^f e_i^k ,
\label{ekav}
\end{equation}
 with $e_i$ denoting the charge of the quark  of the flavour $i$ and $f$ being equal to the number of active
flavours.\\

It is also convenient to 'unfold' the Sudakov form-factor(s) so that the virtual corrections
and real emission terms appear on equal footing in the kernels of the corresponding
system of integral equations.
The unfolded system of CCFM equations in the single loop approximation
takes the following form:

\begin{gather}
f_{NS}(x,Q_t,Q)={\alpha_{em}\over 2\pi}{k_{NS}^0(x)\over Q_t^2} +
f^0_{NS}(x,Q_t) +
\nonumber\\*[5mm]
+\int_0^1dz\int{d^2q\over \pi q^2} {\alpha_s(q^2)\over
2\pi}\Theta(q^2-q_0^2)\Theta(Q-q) P_{qq}(z)\times
\nonumber\\*[5mm]
\times\left[\Theta(z-x)f_{NS}\left({x\over
z},Q^{\prime}_t,q\right)-f_{NS}(x,Q_t,q) \right] ,
\label{ccfmns}
\end{gather}

\begin{gather}
f_{S}(x,Q_t,Q)={\alpha_{em}\over 2\pi}{k_{S}^0(x)\over Q_t^2} +
f_S^0(x,Q_t)+ \nonumber\\*[5mm] + \int_0^1dz\int{d^2q\over \pi
q^2} {\alpha_s(q^2)\over 2\pi}\Theta(q^2-q_0^2)\Theta(Q-q)\times
\nonumber\\*[5mm]
\times\bigg\{\Theta(z-x)\left[P_{qq}(z)f_{S}\left({x\over
z},Q^{\prime}_t,q\right)+ P_{qg}(z)f_{g}\left({x\over
z},Q^{\prime}_t,q\right)\right] -\bigg.
\nonumber\\*[5mm]
\bigg. -P_{qq}(z)f_{S}(x,Q_t,q)\bigg\}, \label{ccfms}
\end{gather}

\begin{gather}
f_{g}(x,Q_t,Q)=f_g^0(x,Q_t)+
\nonumber\\*[5mm]
+ \int_0^1dz\int{d^2q\over \pi q^2} {\alpha_s(q^2)\over
2\pi}\Theta(q^2-q_0^2)\Theta(Q-q)\times
\nonumber\\*[5mm]
\times\bigg\{\Theta(z-x)\left[P_{gq}(z)f_{S}\left({x\over
z},Q^{\prime}_t,q\right)+ P_{gg}(z)f_{g}\left({x\over
z},Q^{\prime}_t,q\right)\right]-\bigg.
\nonumber\\*[5mm]
\bigg.- \big[zP_{gg}(z)+zP_{qg}(z)\big]f_{g}(x,Q_t,q)\bigg\} ,
\label{ccfmg}
\end{gather}

where
\begin{equation}
{\bf Q^{\prime}_t} ={\bf Q_t} + (1-z){\bf q} .
\label{qprimet}
\end{equation}

The functions $k_{NS}^0(x)$ and $k_S^0(x)$ are defined as below:
\begin{equation}
k_{NS}^0(x)=2N_cf(<e^4>-<e^2>^2)[x^2+(1-x)^2] ,
\label{kns}
\end{equation}

\begin{equation}
k_{S}^0(x)=2N_cf<e^2>[x^2+(1-x)^2] ,
\label{kns}
\end{equation}
with $N_c$ denoting the number of colours.
The inhomogeneous terms proportional to $k_{NS}^0(x)$ and
$k_S^0(x)$
in equations (\ref{ccfmns}) and (\ref{ccfms}) respectively reflect the point coupling
of the photon to quarks and antiquarks.
 The functions
$f_{NS}^0(x,Q_t),f_{S}^0(x,Q_t),f_{g}^0(x,Q_t)$
denote the non-perturbative 'hadronic' components of the unintegrated non-singlet, singlet and gluon
distributions respectively.  The parameter $q_0$ is the infrared cut-off.
The splitting functions $P_{ab}(z)$ are the LO splitting functions, i.e.:

$$
P_{qq}(z)={4\over 3} {1+z^2\over 1-z} ,
$$

$$
P_{qg}(z)=f[z^2+(1-z)^2] ,
$$

$$
P_{gq}(z)={4\over 3} {1+(1-z)^2\over z} ,
$$

\begin{equation}
P_{gg}(z)=2N_c\left[{z\over 1-z}+{1-z\over z} + z(1-z)\right] .
\label{splitf}
\end{equation}

\section{CCFM equation in the transverse coordinate reprecentation}
It can easily be observed that the system of CCFM equations in the single
loop approximation (\ref{ccfmns}) - (\ref{ccfmg}) can be diagonalised  by the
Fourier-Bessel transform \cite{JK}:
\begin{equation}
f_{k}(x,Q_t,Q)=\int_0^{\infty}db b J_0(Q_tb)\bar f_{k}(x,b,Q) ,
\label{fb1}
\end{equation}

\begin{equation}
\bar f_{k}(x,b,Q)=\int_0^{\infty}dQ_t Q_t J_0(Q_tb) f_{k}(x,Q_t,Q) ,
\label{fb2}
\end{equation}
where $k=NS,S,g$ and $J_0(u)$ is the Bessel function.  The corresponding system of 
CCFM equations for $\bar f_{NS}(x,b,Q),
\bar f_{S}(x,b,Q)$ and $\bar f_{g}(x,b,Q)$
which follows from equations (\ref{ccfmns}) - (\ref{ccfmg}) reads:

\begin{gather}
\bar f_{NS}(x,b,Q)={\alpha_{em}\over 2\pi}k_{NS}^0(x)\bar f_{pt}^0(b,Q) +
\bar f^0_{NS}(x,b)+
\nonumber\\*[5mm]
+ \int_0^1dz\int{dq^2\over q^2} {\alpha_s(q^2)\over
2\pi}\Theta(q^2-q_0^2)\Theta(Q-q) P_{qq}(z)\times
\nonumber\\*[5mm]
\times\left[\Theta(z-x)J_0\big[(1-z)qb\big]\bar f_{NS}
\left({x\over z},b,q\right)-\bar f_{NS}(x,b,q)
\right] ,
\label{ccfmnsb}
\end{gather}

\begin{gather}
f_{S}(x,Q_t,Q)={\alpha_{em}\over 2\pi}k_{S}^0(x)\bar f_{pt}^0(b,Q) +
\bar f_S^0(x,b)+
\nonumber\\*[5mm]
+ \int_0^1dz\int{dq^2\over q^2} {\alpha_s(q^2)\over 2\pi}
\Theta(q^2-q_0^2)\Theta(Q-q)\times
\nonumber\\*[5mm]
\times\bigg\{\Theta(z-x)J_0\big[(1-z)qb\big]\left[P_{qq}(z)\bar
f_{S}\left({x\over z},b,q\right)+ P_{qg}(z)\bar f_{g}\left({x\over
z},b,q\right)\right] -\bigg.
\nonumber\\*[5mm]
\bigg.- P_{qq}(z)\bar
f_{S}(x,b,q)\bigg\} ,
\label{ccfmsb}
\end{gather}

\begin{gather}
\bar f_{g}(x,Q_t,Q)=\bar f_g^0(x,b)+
\nonumber\\*[5mm]
+ \int_0^1dz\int{dq^2\over q^2}
{\alpha_s(q^2)\over 2\pi}\Theta(q^2-q_0^2)
\Theta(Q-q)\times
\nonumber\\*[5mm]
\times\bigg\{\Theta(z-x)J_0\big[(1-z)qb\big]\left[P_{gq}(z)\bar f_{S}\left({x\over z},b,q\right)+
P_{gg}(z)\bar f_{g}\left({x\over z},b,q\right)\right] -\bigg.
\nonumber\\*[5mm]
\bigg.- \big[zP_{gg}(z)+zP_{qg}(z)\big]\bar f_{g}(x,b,q)\bigg\} .
\label{ccfmgb}
\end{gather}

The  function $\bar f_{pt}^0(b,Q)$ controlling the inhomogeneous term originating from the point-like
interaction  is defined as:

\begin{equation}
\bar f_{pt}^0(b,Q)=\int_{q_0}^Q dQ_t Q_t{J_0(bQ_t)\over Q_t^2} .
\label{barpt}
\end{equation}
In the definition of the inhomogeneous term corresponding to the point interaction of the
photon we have introduced upper limit cut-off equal to $Q$
in the integration over $dQ_t$ in equation (\ref{barpt}).  This is necessary for making
the CCFM formalism compatible with the DGLAP evolution for the integrated
parton distributions $f^{int}_i(x,Q^2)$
\begin{equation}
xf^{int}_i(x,Q^2)=\int_0^{\infty} dQ_t^2 f_i(x,Q_t,Q) .
\label{intun}
\end{equation}
The integrated distributions  $f^{int}_i(x,Q^2)$ are given by  the distributions
$\bar f_i(x,b,Q)$ at $b=0$ i.e.
\begin{equation}
xf^{int}_i(x,Q^2)= 2 \bar f_i(x,b=0,Q) .
\label{intb}
\end{equation}

Equations (\ref{ccfmnsb}) - (\ref{ccfmgb}) are equivalent to the following system
of inhomogeneous differential equations:

\begin{gather}
Q^2{\partial \bar f_{NS}(x,b,Q)\over \partial Q^2}
={\alpha_{em}\over 2\pi}k_{NS}^0(x) {J_0(bQ)\over 2}
+ {\alpha_s(Q^2)\over 2\pi}
\int_0^1dz
P_{qq}(z)\times
\nonumber\\*[5mm]
\times\left[\Theta(z-x)J_0\big[(1-z)Qb\big]\bar f_{NS}\left({x\over z},b,Q\right)
- \bar f_{NS}(x,b,Q)
\right] ,
\label{dccfmnsb}
\end{gather}

\begin{gather}
Q^2{\partial f_{S}(x,Q_t,Q)\over \partial Q^2}=
{\alpha_{em}\over 2\pi}k_{S}^0(x){J_0(bQ)\over 2} +
{\alpha_s(Q^2)\over 2\pi}
\int_0^1dz
\bigg\{\Theta(z-x)J_0\big[(1-z)Qb\big]\times\bigg.
\nonumber\\*[5mm]
\bigg.\times\left[P_{qq}(z)\bar f_{S}\left({x\over z},b,Q\right)+
P_{qg}(z)\bar f_{g}\left({x\over z},b,Q\right)\right] - P_{qq}(z)\bar f_{S}(x,b,Q)\bigg\}
\label{dccfmsb}
\end{gather}

\begin{gather}
Q^2 { \partial \bar f_{g}(x,Q_t,Q)\over \partial Q^2}=
{\alpha_s(Q^2)\over 2\pi}\int_0^1dz
\bigg\{\Theta(z-x)J_0\big[(1-z)Qb\big]\times \bigg.
\nonumber\\*[5mm]
\bigg.\times\left[P_{gq}(z)\bar f_{S}\left({x\over z},b,Q\right)+
P_{gg}(z)\bar f_{g}\left({x\over z},b,Q\right)\right]-
\big[zP_{gg}(z)+zP_{qg}(z)\big]\bar f_{g}(x,b,Q)\bigg\} ,
\label{dccfmgb}
\end{gather}
with the initial conditions:
\begin{equation}
\bar f_i(x,b,q_0)=\bar f_i^0(x,b) ,
\label{bcond}
\end{equation}
where $i$ corresponds to $NS$, $S$ and $g$. In complete analogy to the integrated
parton distributions in a photon  we can  introduce  conventional
decomposition of the distributions $\bar f_i(x,b,Q)$ into their
point-like $\bar f_i^p(x,b,Q)$ and hadronic $\bar f_i^h(x,b,Q)$ components  i.e.
\begin{equation}
\bar f_i(x,b,Q)=\bar f_i^p(x,b,Q) +\bar f_i^h(x,b,Q) .
\label{decomp}
\end{equation}

The point-like components $\bar f_i^p(x,b,Q)$
are the solutions of  inhomogeneous equations (\ref{dccfmnsb}) - (\ref{dccfmgb}) with
the initial conditions
\begin{equation}
\bar f_i(x,b,q_0)=0 .
\label{bcpoint}
\end{equation}
The hadronic components $\bar f_i^h(x,b,Q)$ are the solutions of the homogeneous
 equations  corresponding to equations (\ref{dccfmnsb}) - (\ref{dccfmgb}) with
inhomogeneous terms set equal to zero.  The initial conditions for the hadronic components are given
by equation (\ref{bcond}).

\section{Numerical results}
In this section we present results of the numerical analysis of the CCFM equation
in the single loop approximation for the gluon distribution in a proton.
To this aim we solved equations (\ref{dccfmsb}) and (\ref{dccfmgb}) following the LO DGLAP
analysis performed at \cite{GRS}.
The unintegrated gluon distributions are then calculated from
equation (\ref{fb1}).
 We have assumed the following initial
conditions for the distributions $f_{S}(x,b,Q)$ and $f_{g}(x,b,Q)$
at $Q=q_0$,  where $q_0^2=0.26 GeV^2$ :
\begin{equation}
f_{S}(x,b,q_0)= {1\over 2}x\Sigma(x,q_0^2)F(b) ,
\label{sinput}
\end{equation}
\begin{equation}
f_{g}(x,b,q_0)= {1\over 2}xg(x,q_0^2)F(b) ,
\label{sinput}
\end{equation}
where the form-factor $F(b)$ was assumed to have the the following form
\begin{equation}
F(b)=exp\left(-{b^2\over b_0^2} \right) ,
\label{fb}
\end{equation}
with $b_0^2=4GeV^{-2}$. The functions $\Sigma(x,q_0^2)$ and
$g(x,q_0^2)$, which  are the integrated singlet and gluon
distributions in the photon at the reference scale were taken from refs.
\cite{GRS} and \cite{PION}.  To be precise the
parton distributions in a photon  at the reference scale $Q=q_0$ were obtained in
\cite{GRS} from the VMD model with the parton distributions in
vector mesons assumed to be given by those in a pion and taken
from \cite{PION}.  The singlet and gluon distributions in the
photon at $Q^2=q_0^2$ are
expressed in the following way in terms of the corresponding
distributions in the pion:
\begin{equation}
x\Sigma(x,q_0^2)=\alpha_{em}(G_{\rho}^2+G_{\omega}^2)\big[xq_{v}^{\pi}(x,q_0^2)+
4x\bar q^{\pi}(x,q_0^2)
\big] ,
\label{xsigma}
\end{equation}

\begin{equation}
xg(x,q_0^2)=\alpha_{em}(G_{\rho}^2+G_{\omega}^2)xg^{\pi}(x,q_0^2) ,
\label{}
\end{equation}
with $G_{\rho}^2=0.5$ and $G_{\omega}^2=0.043$.
 The valence quark, antiquark and gluon distributions in a pion for $Q^2=q_0^2$ were
parametrised as below \cite{PION}:
\begin{equation}
xq_{v}^{\pi}(x,q_0^2)=1.129(1.+0.153\sqrt{x})x^{0.504}(1-x)^{0.349} ,
\label{valpi}
\end{equation}

\begin{equation}
x\bar q^{\pi}(x,q_0^2)=0.522(1.-3.243\sqrt{x}+5.206x)x^{0.16}(1-x)^{5.2}
, \label{barqpi}
\end{equation}

\begin{equation}
xg^{\pi}(x,q_0^2)=7.326(1-1.919\sqrt{x}+1.524x)
x^{1.433}(1-x)^{1.326} .
\label{glupi}
\end{equation}

\begin{figure}
\begin{center}
\includegraphics*[angle=0,width=0.6\textwidth]{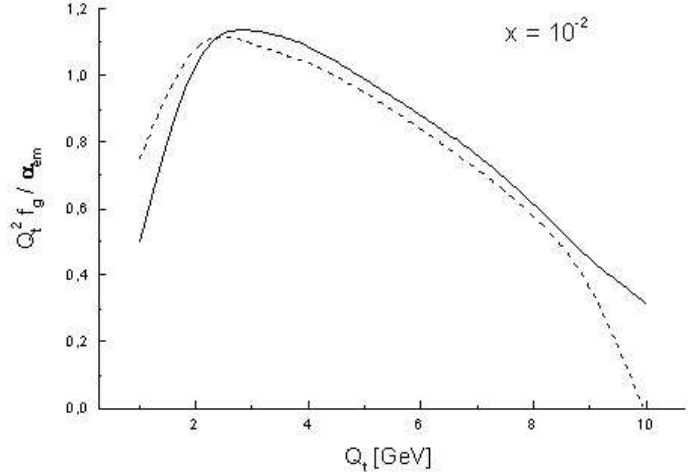}
\caption{The function $Q_t^2 f_g(x,Q_t,Q)/\alpha_{em}$, where
$f_g(x,Q_t,Q)$ is the unintegrated gluon distribution in a photon
plotted as the function of the transverse momentum $Q_t$ of the
gluon for $x=0.01$ and $Q=10 GeV$.  The solid and dashed lines
correspond to the exact solution of the system of the CCFM
equations in the single loop approximation and to the approximate
expression (\ref{ufgm}) respectively.}
\end{center}
\end{figure}

\begin{figure}
\begin{center}
\includegraphics*[angle=0,width=0.6\textwidth]{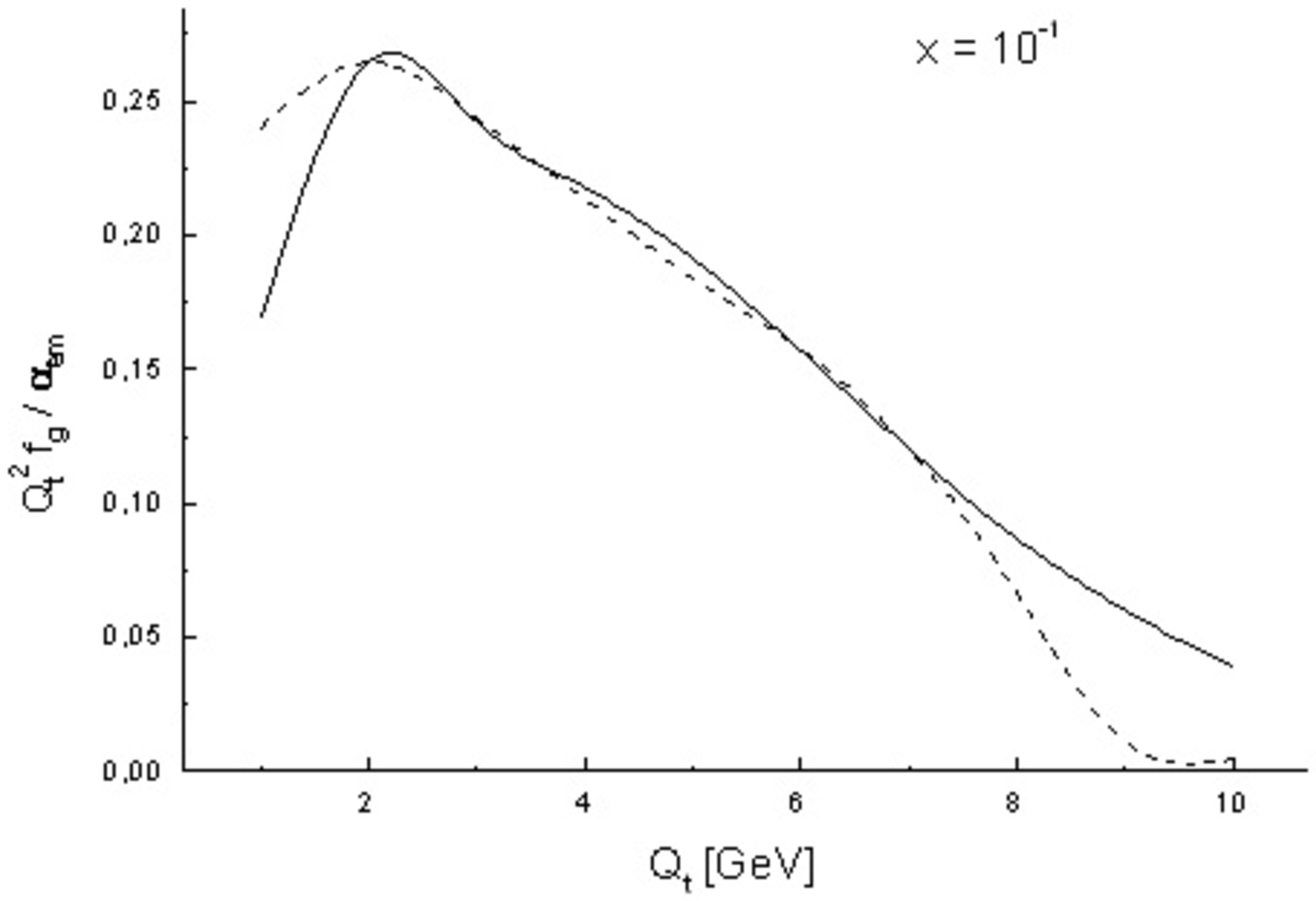}
\caption{The function $Q_t^2 f_g(x,Q_t,Q)/\alpha_{em}$, where
$f_g(x,Q_t,Q)$ is the unintegrated gluon distribution in a photon
plotted as the function of the transverse momentum $Q_t$ of the
gluon for $x=0.1$ and $Q=10 GeV$.  The solid and dashed lines
correspond to the exact solution of the system of the CCFM
equations in the single loop approximation and to the approximate
expression (\ref{ufgm}) respectively.}
\end{center}
\end{figure}

\begin{figure}
\begin{center}
\includegraphics*[angle=0,width=0.6\textwidth]{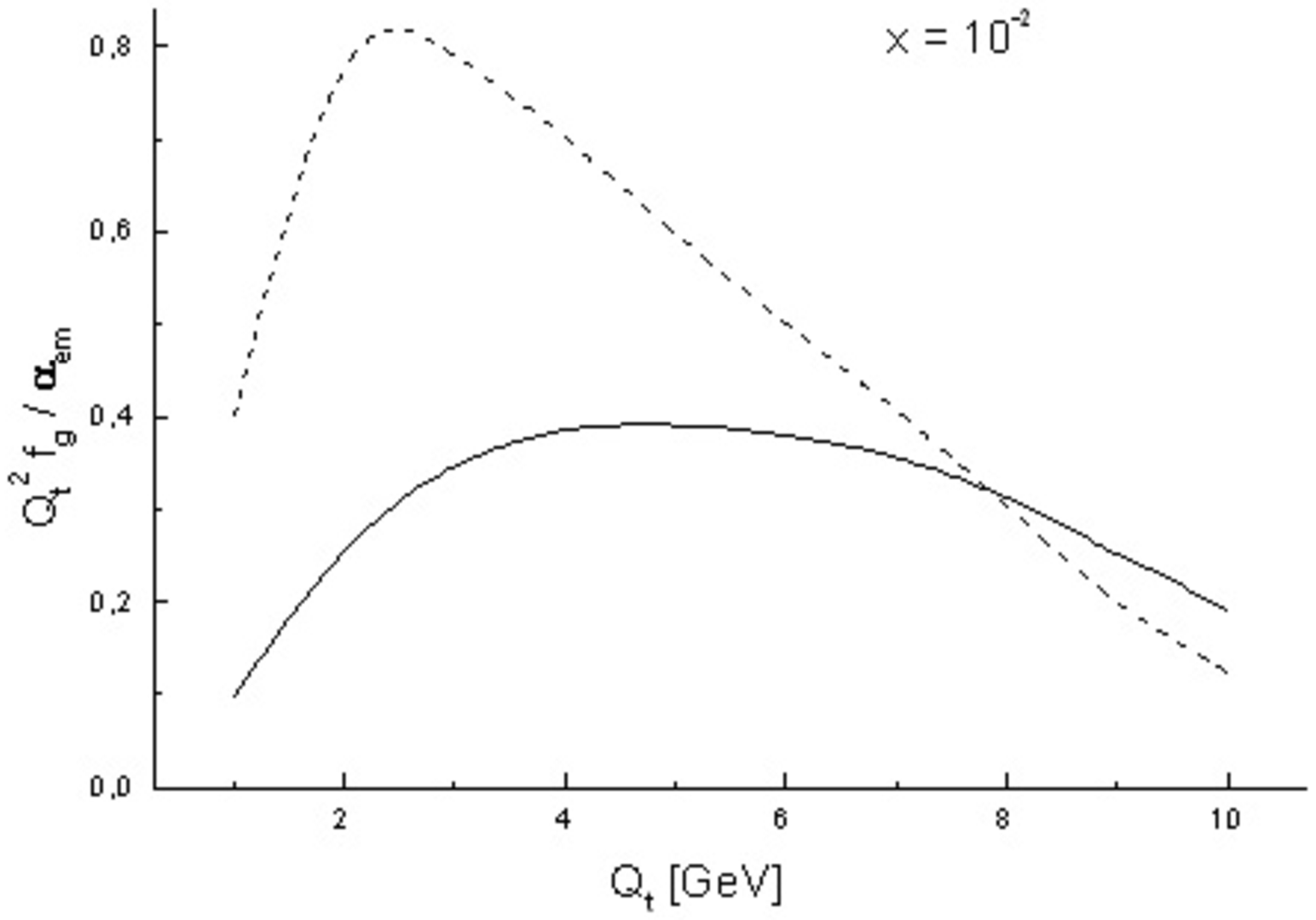}
\caption{The  point-like (solid line) and hadronic (dashed line)
components of the unintegrated gluon distribution in a photon
plotted as functions of the transverse momentum $Q_t$ of the gluon
for $x=0.01$ and $Q=10 GeV$.}
\end{center}
\end{figure}

\begin{figure}
\begin{center}
\includegraphics*[angle=0,width=0.6\textwidth]{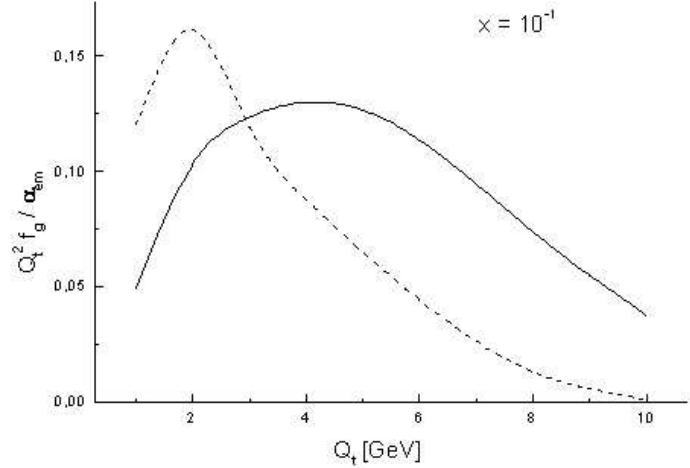}
\caption{The  point-like (solid line) and hadronic (dashed line)
components of the unintegrated gluon distribution in a photon
plotted as functions of the transverse momentum $Q_t$ of the gluon
for $x=0.1$ and $Q=10 GeV$.}
\end{center}
\end{figure}

Results of our calculations concerning unintegrated gluon  distributions in the photon
are presented in
Figures 1 and  2.  We plot in these figures $Q_t^2f_g(x,Q_t,Q)/\alpha_{em}$ as the function
of $Q_t$ at $Q=10GeV$ for two values of $x$, i.e. for $x=0.01$ (Fig. 1)  and $x=0.1$
(Fig. 2).   We compare our result with the approximate expression for
$Q_t^2f_{g}(x,Q_t,Q)/\alpha_{em}$:
 \begin{gather}
Q_t^2f_g(x,Q_t,Q)\simeq
\nonumber\\*[5mm]
\simeq {\alpha_s(Q_t^2)T_g(Q_t,Q)\over 2\pi \alpha_{em}}\int_x^{1-Q_t/Q} dz
\left[P_{gg}(z){x\over z} g\left({x\over z},Q_t^2\right)+
P_{gq}(z){x\over z} \Sigma\left({x\over z},Q_t^2\right)\right] ,
\label{ufgm}
\end{gather}
where the Sudakov-like form-factor is given by:
\begin{gather}
T_g(Q_t,Q) =
\nonumber\\*[5mm]
exp\left\{-
\int_{Q_t^2}^{Q^2}{dq^2\over q^2} {\alpha_s(q^2)\over 2\pi}
\int_0^{1-Q_t/q}
dz^{\prime}\big[z^{\prime} P_{gg}(z^{\prime} )+z^{\prime} P_{qg}(z^{\prime} )\big]\right\} .
\label{ttgm}
\end{gather}
Derivation of approximate relation (\ref{ufgm}), which is similar to that discussed in
\cite{KMR1} is given in the Appendix.  We see that the approximate expression (\ref{ufgm})
reproduces reasonably well exact solution of the CCFM equation for unintegrated gluon
distributions in a photon.  In Figures 3 and 4 we show decomposition of the unintegrated
gluon distrtibutions into their hadronic and point-like components.
 The point-like
component is found to become increasingly important in the region of large $Q_t$.
The relative contribution of this component does also increase with increasing $x$.


\section{Summary and conclusions}
We have considered in this paper the system of CCFM equations in the single loop
approximation for the unintegrated parton   distributions in a photon.
We have extended the conventional CCFM formalism by including
 quarks and the complete splitting functions.  We have utilised the fact that the
CCFM equation(s) in the single loop approximation can be diagonalised by the Fourrier-Bessel
transform. We have found that the unintegrated gluon distributions in a photon
obtained from the
exact solution of the system of CCFM equations in the single loop approximation can be well represented by the
approximate expressions connecting the those diatributions with the
integrated (gluon and quark) distributions and the Sudakov-like form-factor.\\

The novel feature of the CCFM equation for the parton distributions in a photon,
when compared with the hadronic case is the presence of the point-like components.
Those components  become increasingly important at  large values of $x$.  They
have also been found to play important role at large values of the trasnverse
momentum $Q_t$ of the gluon for moderately small values of $x$.\\

The unintegrated gluon distributions which describe the $x$ and
$Q_t$ distributions are important quantities are neede in the description of
the processes which are sensitive  to the transverse momentum of
the gluon.  Their knowledge is in particular necessary for the
description of heavy quark production in $\gamma$ $\gamma$
collisions within the $k_t$ factorisation.  Results obtained in our
paper may therefore be used for the
theoretical analysis of this process. \\

\section*{Acknowledgments}
This research was partially supported
by the EU Fourth Framework Programme `Training and Mobility of Researchers',
Network `Quantum Chromodynamics and the Deep Structure of Elementary
Particles', contract FMRX--CT98--0194 and  by the Polish
Committee for Scientific Research (KBN) grants no. 2P03B 05119 and 5P03B 14420.


\section*{Appendix}
Let us make the following approximation:
\begin{equation}
J(u) \simeq \Theta(1-u) .
\label{besappr}
\end{equation}
It is clear that in this approximation solution of  equations
(\ref{ccfmsb},\ref{ccfmgb})
is independent of $b$ for $Q<1/b$, provided we neglect the $b$ dependence of the
'hadronic' input that is justified at small $b$.
>From (\ref{fb2},\ref{besappr}) we also get:
\begin{equation}
f_{k}(x,Q_t,Q) \simeq 2\quad{\partial \bar f_{k}(x,b=1/Q_t,Q)
\over \partial Q_t^2} .
\label{deriv}
\end{equation}

It is useful to rearrange equations (\ref{ccfmsb},\ref{ccfmgb})
as below:

\begin{gather}
\bar f_{S}(x,b,Q)={\alpha_{em}\over 2\pi }k_{S}^0(x)\bar f_{pt}^0(b,Q) +
\bar f_S^0(x,b) +
\nonumber\\*[5mm]
+\int_0^1dz\int{dq^2\over q^2} {\alpha_s(q^2)\over 2\pi}\Theta(q^2-q_0^2)
\Theta(Q-q)\times
\nonumber\\*[5mm]
\times\bigg\{J_0\big[(1-z)qb\big]\Big[\Theta(z-x)\left(P_{qq}(z)\bar f_{S}
\left({x\over z},b,q\right)+P_{qg}(z)\bar f_{g}\left({x\over z},b,q\right)\right) -
\Big.\bigg.
\nonumber\\*[5mm]
\bigg.\Big.- P_{qq}(z)\bar f_{S}(x,b,q)\Big]+
P_{qq}(z)\big(1-J_0[(1-z)qb]\big)\bar f_{S}(x,b,q)\bigg\} ,
\label{ccfmsb1}
\end{gather}

\begin{gather}
\bar f_{g}(x,b,Q)= \bar f_g^0(x,b) + \int_0^1dz\int{dq^2\over q^2} {\alpha_s(q^2)\over 2\pi}
\Theta(q^2-q_0^2)\Theta(Q-q)\times
\nonumber\\*[5mm]
\times\bigg\{J_0\big[(1-z)qb\big]\Big[\Theta(z-x)\left(P_{gg}(z)\bar f_{g}
\left({x\over z},b,q\right)+P_{gq}(z)\bar f_{S}\left({x\over z},b,q\right)\right)-
\Big.\bigg.
\nonumber\\*[5mm]
\bigg.\Big.- \big(zP_{gg}(z)+zP_{qg}(z)\big)\bar f_{g}(x,b,q)\Big] -
\big[zP_{gg}(z)+zP_{qg}(z)\big]\big(1-J_0[(1-z)qb]\big)\bar f_{g}(x,b,q)\bigg\} .
\label{ccfmgb1}
\end{gather}

Differentiating this equation with respect to $\partial Q_t^2$ for $b^2=1/Q_t^2$
and using equations (\ref{besappr},\ref{deriv}) we get:

\begin{gather}
f_{S}(x,Q_t,Q) \simeq {\alpha_{em}\over 2\pi}{k_{S}^0(x)\over Q_t^2}  +  f_S^0(x,Q_t) +
\nonumber\\*[5mm]
+ \int_0^1dz\int{dq^2\over q^2} {\alpha_s(q^2)\over 2\pi}
\Theta(q^2-q_0^2)\Theta(Q-q)\times
\nonumber\\*[5mm]
\times\bigg\{\delta\big(q^2-Q_t^2/(1-z)^2\big){\Theta(z-x)\over 2Q_t^2}\left[P_{qq}(z) \bar
f_{S}\left({x\over z},b,q\right)+ P_{qg}(z)\bar f_{g}\left({x\over
z},b,q\right)\right] -\bigg.
\nonumber\\*[5mm]
\bigg. - P_{qq}(z)\big[1-\Theta(q^2-Q_t^2/(1-z)^2)\big]
f_{S}(x,Q_t,q)\bigg\} ,
\label{ccfmsq1}
\end{gather}

\begin{gather}
f_{g}(x,Q_t,Q)=  f_g^0(x,Q_t) + \int_0^1dz\int{dq^2\over q^2} {\alpha_s(q^2)\over 2\pi}
\Theta(q^2-q_0^2)\Theta(Q-q)\times
\nonumber\\*[5mm]
\times\bigg\{\delta\big(q^2-Q_t^2/(1-z)^2\big){\Theta(z-x)\over 2Q_t^2}\Big[P_{gg}(z)
\bar f_{g}\left({x\over z},b,q\right)+ P_{gq}(z)\bar f_{S}\left({x\over
z},b,q\right)\Big] -\bigg.
\nonumber\\*[5mm]
\bigg. - \big[zP_{gg}(z)+zP_{qg}(z)\big]\big[1-\Theta(Q_t^2/(1-z)^2-q^2)\big]
f_{g}(x,Q_t,q)\bigg\} .
\label{ccfmgq1}
\end{gather}

In equations (\ref{ccfmsq1},{\ref{ccfmgq1}) we have neglected integrals with the integrands
containing the terms like:
\begin{gather}
\Theta [Q_t-(1-z)q]\left[P_{gg}(z)\Theta(z-x)
{\partial \bar f_{g}\left({x\over z},b=1/Q_t^2,q\right)\over \partial Q_t^2} -\right.
\nonumber\\*[5mm]
\left. - (zP_{gg}(z)+zP_{qg}(z)){\partial \bar f_{g}(x,b=1/Q_t^2,q)
\over \partial Q_t^2}\right] ,
\label{neglect}
\end{gather}


Neglecting those terms is justified, since in the region $q<Q_t/(1-z) \sim Q_t$
 $\bar f_g(x,b=1/Q_t^2,q)$
is independent of $b$ and so its derivative with respect to $\partial Q_t^2$
vanishes.   We next identify:
\begin{equation}
2\bar f_{s}\left({x\over z},b=1/Q_t,q=Q_t/(1-z)\right) \simeq
{x\over z}\Sigma\left({x\over z},Q_t^2\right) ,
\label{apprs}
\end{equation}

\begin{equation}
2\bar f_{g}\left({x\over z},b=1/Q_t,q=Q_t/(1-z)\right) \simeq
{x\over z} g\left({x\over z},Q_t^2\right) .
\label{apprg}
\end{equation}

Substituting (\ref{apprs},\ref{apprg}) into equations (\ref{ccfmsq1},{\ref{ccfmgq1})
we get
\begin{gather}
f_{S}(x,Q_t,Q) \simeq {\alpha_{em}\over 2\pi}{k_{S}^0(x))\over Q_t^2} +
f_S^0(x,Q_t)+
\nonumber\\*[5mm]
+ {\alpha_s(Q_t^2)\over 2\pi Q_t^2}\int_x^{1-Q_t/Q} dz
\left[P_{qq}(z){x\over z} \Sigma \left({x\over z},Q_t^2\right)+
P_{qg}(z){x\over z} g\left({x\over z},Q_t^2\right)\right]-
\nonumber\\*[5mm]
- \int_{q_0^2}^{Q^2} {dq^2\over q^2} {\alpha_s(q^2)\over 2\pi}\int_0^1 dz
P_{qq}(z)
\big[1-\Theta(Q_t^2/(1-z)^2-q^2)\big] f_{S}(x,Q_t,q) ,
\label{ccfmsq2}
\end{gather}

\begin{gather}
f_{g}(x,Q_t,Q) \simeq
f_g^0(x,Q_t)+
\nonumber\\*[5mm]
+ {\alpha_s(Q_t^2)\over 2\pi Q_t^2}\int_x^{1-Q_t/Q} dz
\left[P_{gg}(z){x\over z} g\left({x\over z},Q_t^2\right)+
P_{gq}(z){x\over z} \Sigma\left({x\over z},Q_t^2\right)\right]-
\nonumber\\*[5mm]
- \int_{q_0^2}^{Q^2} {dq^2\over q^2} {\alpha_s(q^2)\over 2\pi}\int_0^1 dz
\big[zP_{gg}(z)+zP_{qg}(z)\big]
\big[1-\Theta(Q_t^2/(1-z)^2-q^2)\big] f_{g}(x,Q_t,q) .
\label{ccfmgq2}
\end{gather}


Let us now define the Sudakov-like form factor  $T_g$


$$
T_g(Q_t,Q)=
$$

\begin{equation}
exp\left\{-
\int_{Q_t^2}^{Q^2}{dq^2\over q^2} {\alpha_s(q^2)\over 2\pi}
\int_0^{1-Q_t/q} dz \big[zP_{gg}(z)+zP_{qg}(z)\big]\right\} .
\label{tg}
\end{equation}

>From equation (\ref{ccfmgq2}) we get the following approximate expression for the
unintegrated gluon distribution:
\begin{gather}
f_g(x,Q_t,Q) \simeq T_g(Q_t,Q){\alpha_s(Q_t^2)\over 2\pi Q_t^2}\int_x^{1-Q_t/Q} dzT_g^{-1}
(Q_t,Q_t/(1-z))\times
\nonumber\\*[5mm]
\times\left[P_{gg}(z){x\over z} g\left({x\over z},Q_t^2\right)+
P_{gq}(z){x\over z} \Sigma\left({x\over z},Q_t^2\right)\right] .
\label{ufg}
\end{gather}

Let us finally notice that:
$$
T_g(Q_t,Q)T_g^{-1}
(Q_t,Q_t/(1-z))=
$$
\begin{equation}
exp\left\{-
\int_{Q_t^2/(1-z)^2}^{Q^2}{dq^2\over q^2} {\alpha_s(q^2)\over 2\pi}
\int_0^{1-Q_t/q}
dz^{\prime} \big[z^{\prime} P_{gg}(z^{\prime} )+z^{\prime} P_{qg}(z^{\prime})\big]\right\} .
\label{ttg}
\end{equation}
Replacing the lower integration limit $Q_t^2/(1-z)^2$ by $Q_t^2$
 in the integral in the argument of the exponent
in equation (\ref{ttg}) we get from  equations (\ref{ufg}) and (\ref{ttg})
equation  (\ref{ufgm}) in Section 4.


\begin{thebibliography}{9999}
\bibitem{DDT} Yu.L. Dokshitzer, D.I. Dyakonov and S.I. Troyan, Phys. Rep. {\bf 58} (1980) 269.
\bibitem{KMR1} M.A. Kimber, A.D.Martin and M.G. Ryskin, Eur. Phys. J. {\bf C12} (2000)
655.
\bibitem{KMKS1}M.A. Kimber, A.D.Martin, J. Kwieci\'nski and A.M. Sta\'sto,
Phys. Rev. {\bf D62} (2000) 094006.
\bibitem{KMR2} M.A. Kimber, A.D.Martin and M.G. Ryskin, Phys. Rev. {\bf D63} (2001)
114027.
\bibitem{MR}A.D. Martin and M.G.Ryskin, Phys. Rev. {\bf D64} (2001) 094017.
\bibitem{KHMR} V.A. Khoze, A.D. Martin and M.G. Ryskin, Eur. Phys. J. {\bf C14} (2000) 525;
{\it ibid.} {\bf C19} (2001) 477; Erratum - {\it ibid.} {\bf C20} (2001) 599.
\bibitem{MIU} G. Gustafson, L. L\"onnblad and G. Miu, hep-ph/0206195.
\bibitem{LUNDSMX} B. Andersson et al., hep-ph/0204115 (to appear at Eur.  Phys. J. C).
\bibitem{CCFM} M. Ciafaloni, Nucl. Phys. {\bf B296} (1988) 49;
S. Catani, F. Fiorani and G. Marchesini, Phys. Lett. {\bf B234} (1990) 339;
Nucl. Phys. {\bf B336} (1990) 18.
\bibitem{GM1} G. Marchesini, in Proceedings of the Workshop "QCD at 200 TeV", Erice, Italy,
1990, edited by L. Cifarelli and Yu. L. Dokshitzer, (Plenum Press, New York, 1992), p. 183.
\bibitem{DKTM}Yu.L. Dokshitzer, V.A. Khoze, S.I. Troyan and A.H. Mueller,
Rev. Mod. Phys. {\bf 60} (1988) 373.
\bibitem{BRW} B.R. Webber
Nucl. Phys. B (Proc. Suppl.) {\bf 18C} (1990) 38.
\bibitem{GMBRW} G.Marchesini and B.R. Webber, Nucl. Phys. {\bf B386} (1992) 215;
B.R. Webber in Proceedings of the Workshop "Physics at HERA", DESY, Hamburg, Germany, 1992, edited by W. Buchm\"uller and G. Ingelman
(DESY, Hamburg, 1992).
\bibitem{GM2}G. Marchesini, Nucl.Phys. {\bf B445} (1995) 49.
\bibitem{KMSU}J. Kwieci\'nski, A.D. Martin and P.J. Sutton, Phys. Rev. {\bf D52} (1995)
1445.
\bibitem{CCFMD} G. Bottazzi, G. Marchesini, G.P. Salam and M Scorletti, Nucl. Phys. {\bf
B505} (1997) 366; JHEP {\bf 9812} (2998) 011.
\bibitem{KGBGT}K. Golec-Biernat, L. Goerlich and J. Turnau, Nucl. Phys. {\bf B527} (1998)
289.
\bibitem{GSAL}G.P. Salam, JHEP {\bf 9903} (1999) 009; Nucl. Phys.Proc. Suppl. {\bf 79} (1999)
426.
\bibitem{JUNG} H. Jung, Nucl. Phys. Proc. Suppl. {\bf 79} (1999) 429; Phys. Rev. {\bf D65}
(2002) 034015; Comput. Phys. Commun. {\bf 143} (2002) 100; J. Phys. {\bf G28}
(2002) 971.
%
\bibitem{JUNGS}H. Jung and G.P. Salam, Eur. Phys. J. {\bf C19} (2001) 351.
\bibitem{JK} J. Kwieci\'nski, Acta Phys. Polon. {\bf B33} (2002) 1809.
\bibitem{BASSETTO} A. Bassetto, M. Ciafaloni, G. Marchesini, Nucl. Phys. {\bf B163} (1980)
429.
%
\bibitem{KODAIRA} J. Kodaira, L. Trentadue, Phys. Lett. {\bf B112} (1982) 66.
\bibitem{PT}Y.I.Dokshitzer, D.I.Dyakonov and S.I.Troyan, Phys. Lett.
{\bf B79} (1978) 269;  G. Parisi and R.Petronzio, Nucl. Phys. {\bf B154} (1979) 427;
J. Collins and D. Soper, Phys. Rev. {\bf D16} (1977) 2219;
J. Collins, D. Soper, G. Sterman, Nucl. Phys. {\bf B250} (1985) 199.
%
\bibitem{SZCZUREK} A. Szczurek, hep-ph/0203050.
\bibitem{GRS} M. Gl\"uck, E. Reya, I. Schienbein, Phys. Rev. {\bf D60} (1999) 054019;
Erratum: {\it ibid.} - {\bf D62} (2001) 019902(E).
\bibitem{PION}M. Gl\"uck, E. Reya, I. Schienbein, Eur. Phys. J. {\bf C10} (1999) 313.
\end{thebibliography}
\end{document}